\documentclass[seceq,preprint]{ptptex}

\usepackage{graphicx}
\preprintnumber[3cm]{TNCT-0401}%<-- [..]: optional width of preprint # column.
\title{Space-time evolution of a condensate \\
     in the interaction representation
       }

\author{  Shinji {\sc Maedan}$^{1,}$\footnote{\tt maedan@tokyo-ct.ac.jp}
    }

\inst{ $^1$ Department of Physics, Tokyo National College of Technology\\
 Tokyo 193-0997, Japan
      }

\abst{The space-time evolution of an inhomogeneous as well as homogeneous condensate
   is studied.
We use the  $ \lambda \phi^4 $ model and adopt the interaction representation in
   which the particle picture of the field $\phi$ with mass $m$ is definitely obtained.
As the initial condition, a coherent state is assumed.
In the case of the homogeneous condensate, the amplitude of it grows continuously
   with time and the perturbative calculation breaks down for large time.
On the other hand, in the case of the inhomogeneous condensate with a finite
   initial size $\sim R$, the behavior of it is so different from that of the 
   homogeneous one.
For $0< t < m R^2/2 $, the growth of the amplitude of the inhomogeneous condensate
   at the origin $ ({\boldsymbol x}={\bf 0}) $ is slowed down compared with
   that of the homogeneous one,
   and the contribution of the interaction term $ (\lambda /4) :\phi^4: $ is
   restricted to the small region around the origin
   of the condensate.
For the late time, $ t > m R^2/2 $, the condensate spreads into space.
  }

\begin{document}

\maketitle

\section{Introduction}
The space-time evolution of a condensate is of considerable interest in the fields of the
   Bose-Einstein condensation in atomic gases, the heating of the universe,
   and the disoriented chiral condensate (DCC).
Here, we mainly focus on the problem of the DCC.
   \cite{rf:AnsRys,rf:BlaKrz,rf:Bjo,rf:KowTay,rf:RajWil395,rf:RajWil577}
In order to study the dynamics of the quantum system having a nonzero condensate,
   the Hartree approximation has been widely used.
In that approximation, the field is split into a mean field (condensate) and a fluctuation,
   and the equation of motion for the mean field and that for
   the fluctuation are obtained.
Then, one needs to solve these coupled equations of motion self-consistently,
   which requires enormous numerical calculations in the
   ( 1+3 ) dimensional case.
So as to simplify the problem, the case with the homogeneous condensate has so 
   far been considered mostly. 
   \cite{rf:BoyVegHol,rf:BoyVegHolLeeSin,rf:CooHabKluMot,rf:TsuVauMat}
However, the DCC has a finite size if exists.
When the size of the condensate is not infinite but finite, how the
   space-time evolution of the condensate or the quantum particle generation
   from it becomes ?

Recently, several authors deal with the inhomogeneous condensate by use of the 
   Hartree approximation.
The condensate is assumed to be inhomogeneous, while it is hard to derive
   the particle picture associated with the quantum fluctuation in the framework
   of the Hartree approximation.
   \cite{rf:AarSmi,rf:SalSmiVin,rf:BraNav,rf:BetPaoSan,
          rf:ArtAhrBaiBerSer,rf:CooDawMih,rf:IkeAsaTsu}
The reason is the following.
As stated above, in the Hartree approximation, the equation of motion for the 
   fluctuation contains the inhomogeneous condensation that depends on the 
   space coordinates as well as the time.
Accordingly, in the mode expansion of the solution to the equation of motion
   for the quantum fluctuation, the physical interpretation of the coefficient
   operator is obscure.\cite{rf:AarSmi}
Although it is difficult to obtain the particle picture in the Hartree approximation,
   the case of the inhomogeneous condensate must be investigated.

In this paper, we use the interaction representation in which the particle
   picture is very clear, and study the space-time evolution of an inhomogeneous
   as well as homogeneous condensate.
In the interaction representation, the field operator $\phi ({\boldsymbol x}, t ) $ obeys
   a free equation of motion,$ ( \Box + m^2 )\phi ({\boldsymbol x}, t ) =0 $, and the
   physical state $ \vert \Psi (t) \rangle$ develops according to the interaction Hamiltonian
   $ H_{\rm I}$,
  $ i ( {d}/{d t} ) \vert \Psi (t) \rangle = H_{\rm I}  \vert \Psi (t) \rangle $.
The field  $\phi ({\boldsymbol x}, t ) $ can be expanded by plane waves 
\begin{equation}
   \phi( {\boldsymbol x}, t) 
 = \int \frac{d^3 {\boldsymbol k}}{ \sqrt{ (2 \pi)^3 2 \, \omega(k) } } \,
    \left\{ a(k) e^{ikx} +  a^{\dag}(k) e^{-ikx} \right\},
  \label{aa}
\end{equation}
where the coefficient operator $ a^{\dag}(k) $ is the creation operator for a boson
   particle with momentum ${\boldsymbol k} $, mass $m$, and energy 
    $ \omega = \sqrt{ {\boldsymbol k}^2 + m^2} $.
A formal solution to the differential equation for $ \vert \Psi (t) \rangle$ is
\begin{eqnarray}
   \vert \Psi (t) \rangle &=& U(t, t_0 ) \vert \Psi (t_0) \rangle  \nonumber  \\
   &=& \vert \Psi (t_0) \rangle 
      -i \int_{t_0}^{t} d t' d {\boldsymbol x'} \, {\cal H}_{\rm I}({\boldsymbol x'}, t') \,
       \vert \Psi (t_0) \rangle  + \cdots.
  \label{ab}
\end{eqnarray}
In actual calculations, the interaction Hamiltonian density ${\cal H}_{\rm I}$ should be treated
   as the perturbative part.
The $ \lambda \phi^4 $ model $(\lambda < 1)$ in $(1+3) $ dimension is used in our analysis
   and the calculations are carried out up to the order ${\cal O} (\lambda)$.
As the sigma model or the Nambu-Jona-Lasinio (NJL) model is regarded
   an effective theory,
   we shall also regard here
   $ \lambda \phi^4 $ model as an effective one.
We make an assumption that a condensate is formed at the initial time  $ t_0 =0 $,
   and examine how it develops.
As the initial state condition, we set up a coherent state that is nonequilibrium
   and study its time development until some finite time.
The behavior of the system after a long time is not discussed in this paper.
We first deal with a homogeneous condensate case, in which the condensate grows in
   proportion to the time and it implies that the perturbative calculation does not work
   for large time.
Boyanovsky et al. show that the homogeneous mean field (condensate) grows with time in
   the $ \lambda \phi^4 $ theory by use of another method than the interaction representation,
   and they also conclude that the perturbative method breaks down for large time.
   \cite{rf:BoyVegHolLeeSin}
Now, if the condensate is taken to have a finite size, how the space-time evolution
   of that condensate becomes compared with the homogeneous case ?

The paper is organized as follows.
In the next section, we calculate the time development of the homogeneous condensate.
The calculation up to the order ${\cal O} (\lambda)$ shows that the condensate grows with time.
In the section 3 and section 4, the space-time evolution of the inhomogeneous condensate is
   studied by comparison with the homogeneous case.
Specially, in the section 3, the free theory $(\lambda =0)$ case is considered.
The section 5 is devoted to the conclusion, and the explicit expression of the Lorentz
   invariant function  $\triangle (\vert {\boldsymbol x} \vert, t )$ is given in the Appendix.
\section{Homogeneous condensate}
The Hamiltonian density ${\cal H}$ of the  $ \lambda \phi^4 $ theory is separated into
   free part  ${\cal H}_0$ and interacting part  ${\cal H}_{\rm I}$,
\begin{eqnarray}
  {\cal H}_0 &=& {1\over2} \, \dot \phi (x) ^2
            +{1\over2} \, (  \nabla \phi (x) )^2 + {m^2 \over 2}  \phi (x)^2,  \label{baa}  \\
  {\cal H}_{\rm I} &=& {\lambda\over4}  : \phi (x)^4 :,
  \label{bab}
\end{eqnarray}
where $ \phi (x) $ is a real scalar field with mass $m$ and $ 0<\lambda < 1 $.
The expectation value of the field $\phi$ (condensate) up to the order ${\cal O}(\lambda)$ is
   given by
\begin{eqnarray}
  & & \langle \Psi (t) \vert \, \phi({\boldsymbol x}, t) \, \vert \Psi (t) \rangle   \nonumber  \\
  &=&  \langle \Psi (0) \vert \, \phi({{\boldsymbol x}}, t) \, \vert \Psi (0) \rangle   \nonumber  \\
  & &  -i \int_{0}^{t} d t'   \int_{-\infty}^{\infty}  d^3 {{\boldsymbol x}'} \, 
         \langle \Psi (0) \vert \,
              [ \, \phi({{\boldsymbol x}}, t) , \, {\cal H}_{\rm I}( {{\boldsymbol x}'}, t') \, ]
             \, \vert \Psi (0) \rangle  + O( \lambda^2 )       \nonumber  \\
  &=&  \langle \Psi (0) \vert \, \phi({{\boldsymbol x}}, t) \, \vert \Psi (0) \rangle   \nonumber  \\
  & &  + \, \lambda \int_{0}^{t} d t'  \int_{-\infty}^{\infty} d^3 {{\boldsymbol x}'} \, 
             \Delta ( \vert {\boldsymbol x} - {\boldsymbol x}' \vert, t-t' )
          \langle \Psi (0) \vert  \, : \phi({{\boldsymbol x}'}, t' )^3 :  \, \vert \Psi (0) \rangle 
       + {\cal O}( \lambda^2 ),             \nonumber  \\
  \label{bb}
\end{eqnarray}
where the initial time $t_0 =0 $ and Eq.(\ref{ab}) has been used.
Since we work in the interaction representation, $ \phi (x) $ obeys a free field equation
\begin{equation}
    ( \Box + m^2 ) \phi({{\boldsymbol x}}, t) =  0,
  \label{bc}
\end{equation}
and the Lorentz invariant function  $  \Delta ( \vert {\boldsymbol x} - {\boldsymbol x}' \vert, t-t' ) $ is defined by
\begin{equation}
 [  \, \phi ({\boldsymbol x}, t ) , \,  \phi( {\boldsymbol x}', t' )  \, ] 
   = i \, \Delta ( \vert {\boldsymbol x} - {\boldsymbol x}' \vert, t-t' ).
  \label{bd}
\end{equation}

At the time $t_0=0$, we set a coherent state \cite{rf:KowTay,rf:IshMarTak} as
  the initial state
\begin{equation}
  \vert \Psi ( 0 ) \rangle 
   = N_D   \exp \left\{ \int d {\boldsymbol k} f ({\boldsymbol k}) \, 
      a^\dagger ({\boldsymbol k}) \right\} \vert 0 \rangle.
  \label{be}
\end{equation}
$N_D$ is a normalization constant and the expectation value of number density is given by
 $ \langle \Psi (0) \vert  \, a^\dag ({\boldsymbol k}) a({\boldsymbol k}) \, \vert \Psi (0) \rangle
    = \vert f({\boldsymbol k})  \vert^2 $. 
The coherent state has the property that it is an eigenstate of the
   annihilation operator,
\begin{equation}
  a ({\boldsymbol k}) \vert \Psi (0) \rangle =  f ({\boldsymbol k}) \vert \Psi (0) \rangle,
  \label{bf}
\end{equation}
then, one has the following relation
\begin{equation}
    \langle \Psi (0) \vert  \, : \phi({{\boldsymbol x}}, t )^3 :  \, \vert \Psi (0) \rangle 
   =  \left\{ \langle \Psi (0) \vert  \, \phi({{\boldsymbol x}}, t )  \, \vert \Psi (0) \rangle \right\}^3.
  \label{bg}
\end{equation}
With the help of this relation, the condensate  (\ref{bb}) becomes
\begin{eqnarray}
  & & \langle \Psi (t) \vert \, \phi({{\boldsymbol x}}, t) \, \vert \Psi (t) \rangle   \nonumber  \\
  &=&  \langle \Psi (0) \vert \, \phi({{\boldsymbol x}}, t) \, \vert \Psi (0) \rangle   \nonumber  \\
  & &  + \, \lambda \int_{0}^{t} d t'  \int_{-\infty}^{\infty} d^3 {{\boldsymbol x}'} \,
          \Delta ( \vert {\boldsymbol x} - {\boldsymbol x}' \vert, t-t' )
         \left\{ \langle \Psi (0) \vert  \, \phi({{\boldsymbol x}'}, t' )  \, \vert \Psi (0) \rangle \right\}^3 
        + {\cal O} ( \lambda^2 ).  \nonumber  \\
  \label{bh}
\end{eqnarray}
Notice that the state $ \vert \Psi (t) \rangle $ with $t>0$ is no longer a coherent state.
We are interested in the problem;
   what is the difference in the space-time evolution between the homogeneous and
   inhomogeneous condensation.

At first, let us study the time evolution of the homogeneous condensation which is
   obtained by including only zero mode $ f({\boldsymbol k}) \propto \delta({\boldsymbol k}) $ in Eq.(\ref{be}).
In the free theory  ($ \lambda=0 $), the time dependent condensate (\ref{bh}) is
\begin{equation}
   \langle \Psi (0) \vert  \, \phi({{\boldsymbol x}}, t )  \, \vert \Psi (0) \rangle
  =  \langle \Psi (0) \vert  \, \phi( {\bf 0}, 0 )  \, \vert \Psi (0) \rangle \cos (m \, t)
  \equiv v  \cos (m \, t),
  \label{bi}
\end{equation}
where the mode expansion  (\ref{aa}) has been used.
This is the simple harmonic oscillator with frequency $m$ and amplitude $v$.
Due to the coherent state assumed at $t_0=0$, such a wave behavior of the condensate emerges.
With the interaction term included, the condensate becomes
\begin{eqnarray}
  & & \langle \Psi (t) \vert \, \phi({{\boldsymbol x}}, t) \, \vert \Psi (t) \rangle   \nonumber  \\
  &=& v \cos (m \, t )  
          +\lambda  \,  v^3 
           \int_{0}^{t} d t'   \cos^3 (m \, t' \,)  \int_{-\infty}^{\infty} d^3 {{\boldsymbol x}'} \, 
          \Delta ( \vert {\boldsymbol x} - {\boldsymbol x}' \vert, t-t' ) 
         + {\cal O} ( \lambda^2 )                                      \nonumber  \\
  &=&   v \cos (m \, t )   
         +  {(-\lambda) \over 16 \, m^2} \, v^3  \sin ( m \, t ) 
          \biggr[ \, 6 \, m \, t + \sin ( 2 m \, t ) \biggr]     
           + {\cal O}( \lambda^2 ),
  \label{bj}
\end{eqnarray}
in which we have used (see Appendix)
\begin{equation}
    \int_{-\infty}^{\infty} d^3 {{\boldsymbol x}'} \, \Delta ( \vert {\bf 0} - {\boldsymbol x}' \vert, t-t' )  
    =  -{1 \over m} \sin \{m (t-t' ) \}.
 \label{bk}
\end{equation}
For the large time $t$, the amplitude of the condensate grows as a linear like function of $t$,
   which implies that the perturbative calculation breaks down for large time.
Boyanovsky et al. have been studied the same  $ \lambda \phi^4 $ theory with the homogeneous condensate
   by another method than the interaction representation. \cite{rf:BoyVegHolLeeSin} 
They also found the mean field (condensate) grows as a function of time,
   and concluded that the perturbation theory breaks down at late times.
If the disoriented chiral condensate exists, it should have a finite size, not an infinite size.
We therefore need to deal with a finite size condensation and study its space-time
   evolution, which will be considered in the next section 3 and section 4.
\section{Inhomogeneous condensate: \\ the case of free theory}
In this section, we investigate an inhomogeneous condensate in the free theory $(\lambda =0)$.
Even if the theory is free, the shape of the condensate with the finite 
   size $\sim R$ changes in a complicated way as the time passes.
So we begin with the case of the free theory.
The case of the interaction theory  ($ 0< \lambda <1 $ ) is studied in the next section 4.

Since a relatively large condensate is of interest, the condensation with the size $ R >> 1/m $ 
   is considered below.
At the initial time $t_0=0$, we set up a coherent state with the finite size $\sim R$
   representing the condensate.
Ishihara, Maruyama, and Takagi proposed the coherent state having the momentum distribution
 \cite{rf:IshMarTak}
\begin{equation}
  f ({\boldsymbol k}) = \sqrt{ \frac{\omega}{2} } \, v \,  R^3 \exp (- R^2 {\boldsymbol k}^2 /2).
  \label{ca}
\end{equation}
By use of this distribution in Eq.(\ref{be}),
   the expectation value of the field $\phi$ at $t=0$ indeed has a finite size $R$,
\begin{eqnarray}
  \langle \Psi (0) \vert  \, \phi({{\boldsymbol x}}, t=0) \, \vert \, \Psi (0) \rangle  
    &=&  \int \frac{d^3 {\boldsymbol k}}{ \sqrt{ (2 \pi)^3 2 \, \omega(k) } } \,
           f(k) \, 2 \cos \, ( {\boldsymbol k} \cdot {\boldsymbol x}  )      \nonumber  \\
    &=& v \,  \exp  \left( - \frac{ r^2 }{2 R^2}  \right),
  \label{cb}
\end{eqnarray}
where $ \omega = \sqrt{ {\boldsymbol k}^2 + m^2} $, $ r= \vert {\boldsymbol x} \vert $, and the condensate is
   spherically symmetric.
In order to localize the condensate, one needs to superpose a lot of modes 
    $ \cos \, ( {\boldsymbol k} \cdot {\boldsymbol x}  )   $.
Hereafter, we study the space-time evolution of this condensate (\ref{cb}) with no
   interaction term.

In the interaction representation, the state does not evolve,
   $ \vert \, \Psi (t) \rangle = \vert \, \Psi (0) \rangle $, when the theory is free ($\lambda=0$).
Then, the condensate depending on the space coordinates and the time is easily obtained,
\begin{eqnarray}
   & &\langle \Psi (0) \vert  \, \phi({{\boldsymbol x}}, t) \, \vert \, \Psi (0) \rangle    \nonumber \\
  &=& \int \frac{d^3 {\boldsymbol k}}{ \sqrt{ (2 \pi)^3 2 \, \omega(k) } } \,
           f(k) \, 2 \cos \{ {\boldsymbol k} \cdot {\boldsymbol x} - \omega (k) \, t \}   \nonumber \\
  &=& \frac{4 \pi v R^3}{ (2 \pi)^{3/2} \,  r } \int_{0}^{\infty} d k \exp (- R^2 k^2/2)
         \, k \, \sin (r k ) \cos \{ t \, \sqrt{k^2 +m^2}  \}.
  \label{cc}
\end{eqnarray}
This condensate is the superposition of many different modes 
   $ \cos \{ {\boldsymbol k} \cdot {\boldsymbol x} - \omega (k) \, t \} $ with the weight $f({\boldsymbol k})$.
Such a wave behavior of the condensate comes from the initial state condition
   which we set up the coherent state.
Because the theory is free, each mode moves independently and does not interact with each other.

We study the space-time evolution of the condensate  (\ref{cc}) using some approximations when
   the time $t$ is comparatively small, $ 0< t < m R^2/2 $.
The expression  (\ref{cc}) is rewritten as 
\begin{eqnarray}
   & & \langle \Psi (0) \vert  \, \phi({{\boldsymbol x}}, t) \, \vert \, \Psi (0) \rangle    \nonumber \\
  &=& \frac{4 \pi v R_m}{ (2 \pi)^{3/2} \,  r_m } \int_{0}^{\infty} d u \exp (- u^2/2)
         \, u \, \sin ( ({ r_m  \over R_m } ) u ) \cos
          \left\{ t_m \, \sqrt{ 1+ { u^2 \over R_m^2 } } \, \right\}, 
  \label{cd}
\end{eqnarray}
where the following dimensionless parameters have been introduced,
\begin{eqnarray}
  & &   t_m \equiv m \, t,  \hskip1cm
         r_m \equiv m \, r,  \hskip1cm  R_m \equiv m \, R,  \hskip1cm   u \equiv kR.
  \label{ce}
\end{eqnarray}
The function  $ \cos  \left\{ t_m \, \sqrt{ 1+ u^2 / R_m^2  } \, \right\} $ 
   in Eq.(\ref{cd}) is approximated as follows.
In the definite integral in Eq.(\ref{cd}) with respect to the integral variable 
   $0 \le u \le \infty $, the contribution of the range $ u>2$ is small because of the factor 
   $ \exp (-u^2/2 ) $.
Under the following conditions,
\begin{equation}
   u < 2 ,   \hskip1cm  R_m >> 1 \, ,    \hskip1cm  0 < \frac{u^2}{2 R_m^2} t_m <1,
  \label{cf}
\end{equation}
we make an approximation with
\begin{eqnarray}
 \cos  \left\{ t_m \, \sqrt{ 1+ { u^2 \over R_m^2 } } \, \right\} 
  & \approx & \cos  \left\{ t_m +  \frac{u^2}{2 R_m^2} t_m  \, \right\}      \nonumber \\
  & \approx &  \left\{ 1- {1 \over 2!} \left(  \frac{u^2}{2 R_m^2} t_m  \right)^2 \right\}  \cos ( t_m ) 
          -  \left\{ {1 \over 1!} \left(  \frac{u^2}{2 R_m^2} t_m  \right) \right\}  \sin(t_m).       \nonumber \\
  \label{cg}
\end{eqnarray}
With this approximation, 
   \footnote{ If one wishes to use this approximation for late times,
   $ 1 < \frac{u^2}{2 R_m^2} t_m $, higher orders in the expansion of 
   $ \sin \left(  \frac{u^2}{2 R_m^2} t_m \right) $ or
   $ \cos  \left(  \frac{u^2}{2 R_m^2} t_m \right) $ should be included.   }
the condensate  (\ref{cd}) is
\begin{eqnarray}
   & & {1 \over v} \langle \Psi (0) \vert  \, \phi({{\boldsymbol x}}, t) \,
                 \vert \, \Psi (0) \rangle    \nonumber \\
 & \approx & \frac{4 \pi  R_m}{ (2 \pi)^{3/2} \,  r_m } \int_{0}^{\infty} d u \exp (- u^2/2)
         \, u \, \sin ( ({ r_m  \over R_m } ) u )        \nonumber \\
 & &    \times \left[ \cos ( t_m ) \times \left\{ 1- {1 \over 2 !} \left(  \frac{u^2}{2 R_m^2} t_m  \right)^2 \right\}
          - \sin(t_m)  \times \left\{ {1 \over 1 !} \left(  \frac{u^2}{2 R_m^2} t_m  \right) \right\}  \right]   \nonumber \\
 &=& \cos(t_m) \exp \left\{ -{1 \over 2} \left( {r_m \over R_m} \right)^2 \right\}
       \left[ 1 - {1 \over 2 !} \left( \frac{t_m}{2 R_m^2} \right)^2
        \left\{ 15 -10 \left( {r_m \over R_m} \right)^2 + \left( {r_m \over R_m} \right)^4 \right\} \right]   \nonumber \\
 & & -\sin(t_m) \exp \left\{ -{1 \over 2} \left( {r_m \over R_m} \right)^2 \right\}
       \left[  {1 \over 1 !} \left( \frac{t_m}{2 R_m^2} \right)
        \left\{ 3 -\left( {r_m \over R_m} \right)^2  \right\} \right],
  \label{ch}
\end{eqnarray}
which holds for $R_m >>1 $ and $   \frac{2}{R_m^2} t_m <1  $.
Moreover, when $R_m >>1 $ and $   \frac{2}{R_m^2} t_m <<1  $, one can approximate as
\begin{equation}
    {1 \over v} \langle \Psi (0) \vert  \, \phi({{\boldsymbol x}}, t) \, \vert \, \Psi (0) \rangle  
   \approx  \exp  \left( - \frac{ r_m^2 }{2 R_m^2}  \right) 
            \cos \,( t_m ),
  \label{ci}
\end{equation}
then, the condensate keeps its initial shape 
   $  \exp  \left( - \frac{ r_m^2 }{2 R_m^2}  \right)  $ 
   as far as $ t \ll m R^2/2 $.
In the time zone $t \sim  m \, R^2/2 $, the condensate changes its shape gradually.

How does the condensate alter in shape after the time $ t> m R^2/2 $ ?
The condensate spreads and it is no longer localized in the region of size $ R$.
In general, if the size $R$ of the condensate at the initial time becomes smaller,
   more and more modes are needed to localize it.
Therefore, the condensate spreads rapidly with time.
\section{Inhomogeneous condensate: \\ the case of interaction theory}

In this section, we consider the finite size condensate in the interaction theory.
In the free case ($\lambda=0$) discussed in the previous section,
   the shape of the condensate is not so much changed until
    $  t< m \, R^2/2 $.
With the interaction included, we calculate the condensate when  $ 0< t< m \, R^2/2 $.
The effect of the interaction term ${\cal H}_{\rm I}$ on
   the space-time evolution of the finite size condensation will be clarified.

In order to obtain the expectation value
   $ \langle \, \Psi (t) \vert \, \phi({{\boldsymbol x}}, t) \, \vert \Psi (t) \, \rangle $
   including the interaction effects, one needs to know the quantity
   $ \langle \Psi (0) \vert \, \phi({{\boldsymbol x}'}, t') \, \vert \Psi (0) \rangle  $,
   $ ( 0 \le t' \le t )$.
Since we already have the approximated expression Eq.(\ref{ch}) or (\ref{ci}) of that quantity,
   we can use them to calculate the condensate within the time $ t < m R^2/2 $.
\subsection{Condensate when $ t \ll m R^2/2 $ }
To start with, we consider the condensate at early time $ t \ll m R^2/2 $ 
   in this subsection during which the approximation (\ref{ci}) can be used.
The condensate (\ref{bh}) then becomes
\begin{eqnarray}
  & & \langle \Psi (t) \vert \, \phi({{\boldsymbol x}}, t) \, \vert \Psi (t) \rangle   \nonumber  \\
  & \approx &  v \exp \left( - \frac{  \vert {\boldsymbol x} \vert^2 }{2 R^2}  \right)  \cos \, (m \, t )   \nonumber  \\
  & &  +   \lambda \, v^3 \int_{0}^{t} d t'   \cos^3 \, (m \, t' )
         \int_{-\infty}^{\infty} d^3 {{\boldsymbol x}'} \, \Delta ( \vert {\boldsymbol x} - {\boldsymbol x}' \vert, t-t' )        
             \exp \left( - \frac{ 3 \vert {\boldsymbol x}' \vert^2 }{2 R^2}  \right)   + O( \lambda^2 ).  \nonumber  \\
  \label{da}
\end{eqnarray}
To find the space-time evolution of Eq.(\ref{da}) concretely, let us first focus on the
   time evolution of the condensate at the origin $ {\boldsymbol x}={\bf 0 }$, 
    $ \langle \, \Psi (t) \vert \, \phi( {\bf 0 }, t) \, \vert \Psi (t) \, \rangle $,
   and later analyze the shape of the condensate 
    $ \langle \, \Psi (t) \vert \, \phi( {{\boldsymbol x} }, t) \, \vert \Psi (t) \, \rangle $
   at each time $ (\, t \ll m R^2/2 \,)$.

At the point $ {\boldsymbol x}={\bf 0 }$, the condensate (\ref{da}) is
\begin{eqnarray}
  & & \langle \Psi (t) \vert \, \phi( {\bf 0 }, t) \, \vert \Psi (t) \rangle   \nonumber  \\
  & \approx &  v  \cos \, (m \, t )   \nonumber  \\
  & &  +   \lambda \, v^3 \int_{0}^{t} d t'   \cos^3 \, (m \, t' )
         \int_{-\infty}^{\infty} d^3 {{\boldsymbol x}'} \, \Delta ( \vert {\bf 0} - {\boldsymbol x}' \vert, t-t' )        
             \exp \left( - \frac{ 3 \vert {\boldsymbol x}' \vert^2 }{2 R^2}  \right)   + O( \lambda^2 ).  \nonumber  \\
  \label{db}
\end{eqnarray}
If the size is infinite $R = \infty $, this expression coincides with the homogeneous case Eq.(\ref{bj}),
   as it should be.
When the initial size $R$ of the inhomogeneous condensate (\ref{db})
   becomes smaller, how the time evolution of it changes compared with the homogeneous
   case Eq.(\ref{bj}) ?
During the time $ 0< t \ll m R^2/2 $, we perform the numerical calculations of Eq.(\ref{db}) 
   with $ R_m = 10^4, 15, 10 $ so as to compare the time evolution of the inhomogeneous
   condensation (\ref{db}) with that of the homogeneous one (\ref{bj}).
The numerical result with $\lambda =0.1$ shows that, during $ 0< t \ll m R^2/2 $,
   the difference between
   the amplitude of the homogeneous condensate (\ref{bj}) and that of the inhomogeneous
   condensate (\ref{db}) with the extremely large size $ R_m = 10^4$ is less than $0.003 v$,
   therefore the condensate with $ R_m = 10^4$ can be regarded as the homogeneous
   condensate ($R_m = \infty $). 
In Fig.1, Fig.2, and Fig.3, the time evolution of the condensate (\ref{db})
   with $ R_m = 10^4, 15, 10 $ are represented,
   from which we can observe that the growth of the condensate's amplitude
   at ${\boldsymbol x}={\bf 0}$ is slowed down when its initial size $R_m$ becomes smaller.
%%%%%%%%%%%%%%%%%%%%%%%%%%%%%%%%%%%%%%%%%%%%%%%

\begin{figure}
   \centerline{\includegraphics[height=8cm] {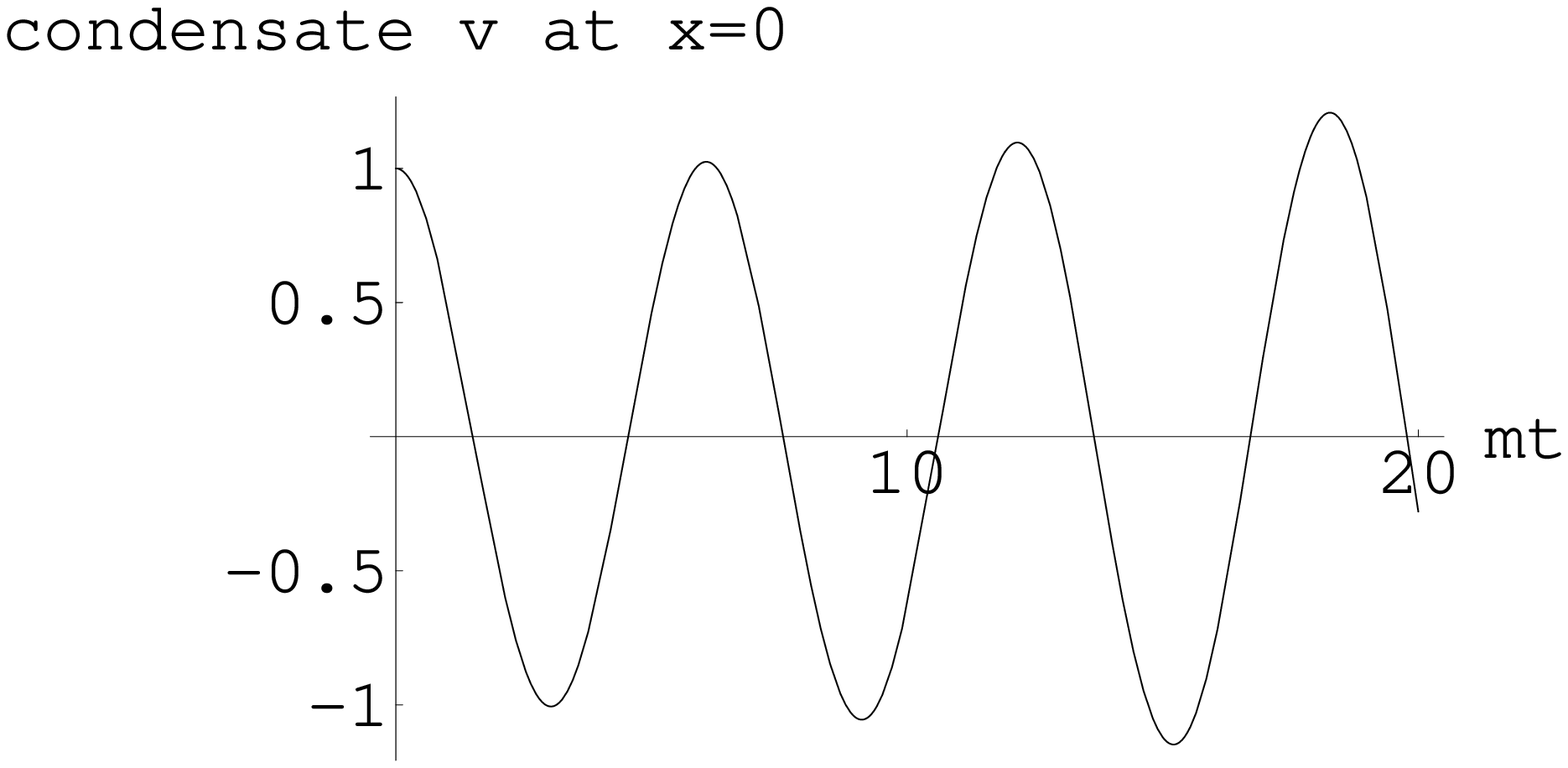}}
   \caption{ Time evolution of the condensate at ${\boldsymbol x}={\bf 0}$ 
             for $R_m =10^4$ when $0<t_m<20$.}
\label{fig:1}
\end{figure}
\begin{figure}
   \centerline{\includegraphics[height=8cm] {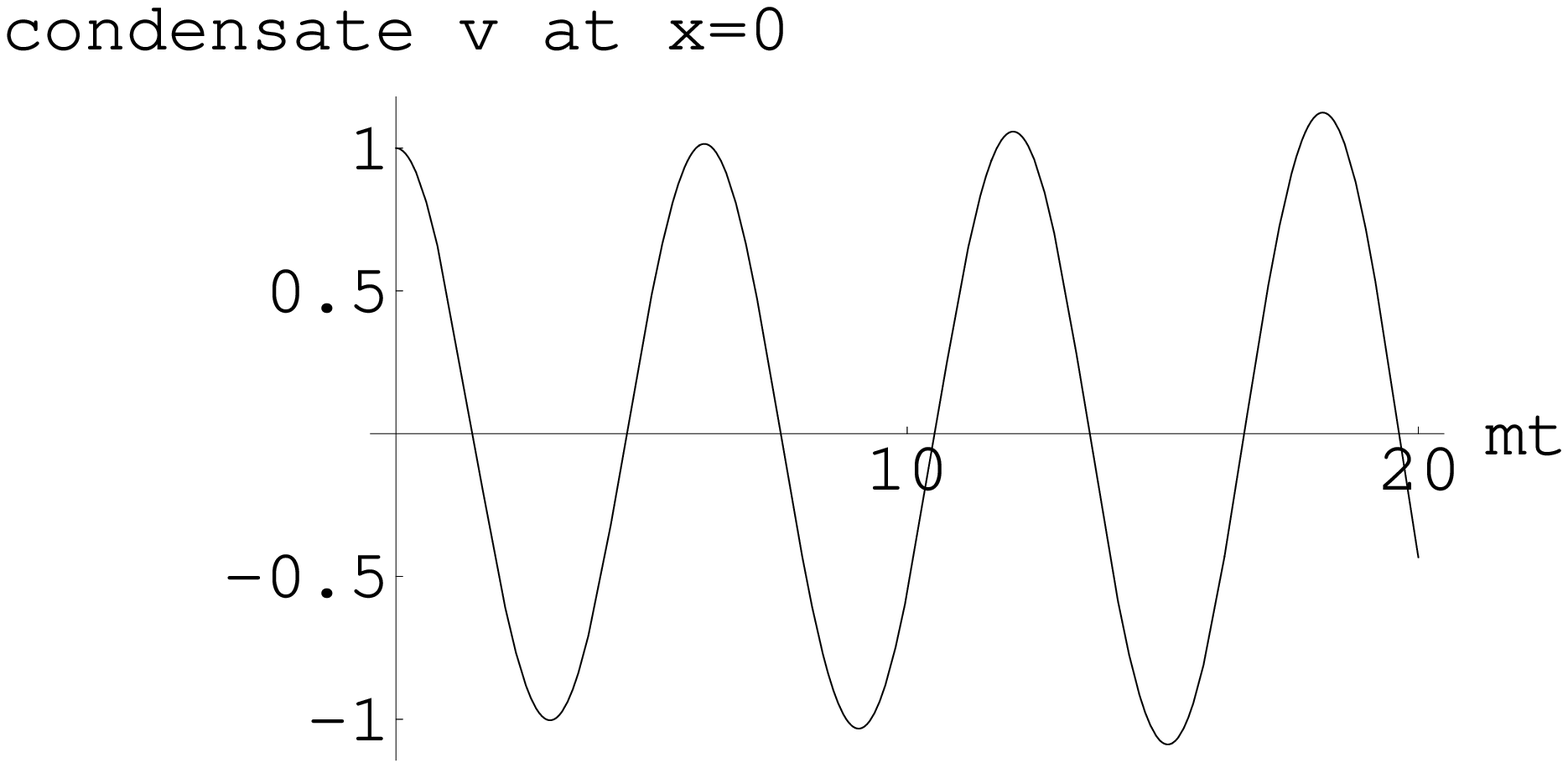}}
   \caption{ Time evolution of the condensate at ${\boldsymbol x}={\bf 0}$ 
             for $R_m =15$ when $0<t_m<20$.}
\label{fig:2}
\end{figure}
\begin{figure}
   \centerline{\includegraphics[height=8cm] {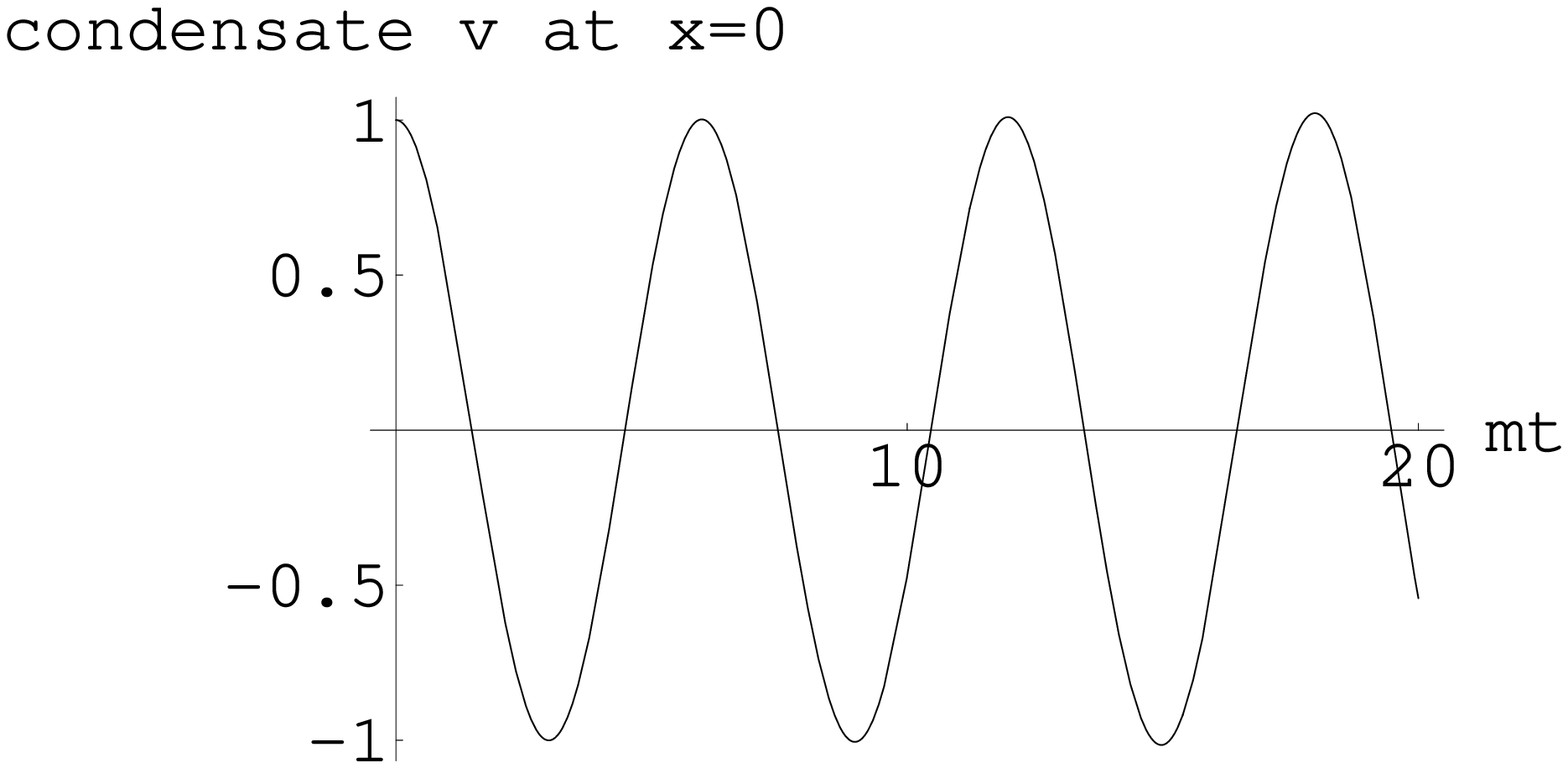}}
   \caption{ Time evolution of the condensate at ${\boldsymbol x}={\bf 0}$ 
             for $R_m =10$ when $0<t_m<20$.}
\label{fig:3}
\end{figure}
%%%%%%%%%%%%%%%%%%%%%%%%%%%%%%%%%%%%%%%%%%%%%%%%%%%%%%%%%%

What is the reason of such a behavior of the condensate ?
The difference between the homogeneous (\ref{bj}) and the inhomogeneous condensate at
   ${\boldsymbol x}={\bf 0}$, (\ref{db}), is a factor 
   $ \exp \left( -  3 \vert {\boldsymbol x}' \vert^2 / 2 R^2  \right)  $  in the integral.
We obtain the following behavior of the integral by carrying out numerical calculation,
\begin{eqnarray}
   & & \int_{-\infty}^{\infty} d^3 {{\boldsymbol x}'} \,
        \Delta ( \vert {\boldsymbol x} - {\boldsymbol x}' \vert, t-t' ) \cdot
          \exp \left( - \frac{ 3 \vert {\boldsymbol x}' \vert^2 }{2 R^2}  \right)  \nonumber  \\
   & & \hskip2.5cm =
     \left\{  \begin{array}{ll}
           -{1 \over m} \sin \{m (t-t' ) \}  \, ,                &  \; ( R = \infty )   \\
           \mbox{  damping oscillation with $( t-t' )$  }  \, ,  &  \;  (\mbox{  finite $R$  } )
                 \end{array}
     \right.
 \label{dc}
\end{eqnarray}
where an analytic integral calculation is possible when $R = \infty $ (see Appendix).
When the size $R$ becomes smaller, the damping in Eq.(\ref{dc}) occurs more rapidly.
Hence, the damping behavior observed in Eq.(\ref{dc}) is one of the reasons why the growth
   of the condensate (at  ${\boldsymbol x}={\bf 0}$ ) is slowed down
   if its initial size $R$ becomes smaller.
This explanation is entirely mathematical, so it is desirable to consider also a physical
   reason, which will be given below.
Let us pay attention to the term of the order ${\cal O}(\lambda)$ in the condensate (\ref{bh}).
The Lorentz invariant function $  \Delta  $ in that term is not zero 
   $  \Delta ( \vert {\boldsymbol x} - {\boldsymbol x}' \vert, t-t' ) \ne 0 $ only for
   $ \vert {\boldsymbol x} - {\boldsymbol x}' \vert^2 \le (t-t' )^2 $ (i.e., time-like interval ).
It represents the causality of  $  \Delta  $.
In other words, in Eq.(\ref{bh}), the expectation value 
    $ \langle \, \Psi (0) \vert  \, \phi({{\boldsymbol x}'}, t') \, \vert \, \Psi (0) \, \rangle $ 
   can contribute to the condensate
   $ \langle \, \Psi (t) \vert  \, \phi({{\boldsymbol x}}, t) \, \vert \, \Psi (t) \rangle $
   if the two space-time points $ ( {\boldsymbol x}', t' ) $ and $  ( {\boldsymbol x}, t )  $
   are connected by
   a time-like interval.
In the case of the homogeneous condensate, its size $R$ is infinite.
Therefore, all the region $ \vert {\boldsymbol x}' \vert \ge 0 $ of 
   $ \langle \, \Psi (0) \vert  \, \phi({{\boldsymbol x}'}, t') \, \vert \, \Psi (0) \, \rangle $ 
   does contribute to the condensate
   $  \langle \,  \Psi (t) \vert  \, \phi({{\boldsymbol x}}, t) \, \vert \, \Psi (t)  \, \rangle $ 
   if $ ( {\boldsymbol x}', t' ) $ and $  ( {\boldsymbol x}, t )  $ are connected by a time-like interval.
On the other hand, in the case of the inhomogeneous condensate with the finite size $R$, only the
   restricted region $ \vert {\boldsymbol x}' \vert < R $ of
   $  \langle \, \Psi (0) \vert  \, \phi({{\boldsymbol x}'}, t') \, \vert \, \Psi (0)  \, \rangle $
   can contribute to the condensate 
   $  \langle \,  \Psi (t) \vert  \, \phi({{\boldsymbol x}}, t) \, \vert \, \Psi (t)  \, \rangle $ 
   even if $ ( {\boldsymbol x}', t' ) $ and $  ( {\boldsymbol x}, t )  $ are connected by
   a time-like interval.
This is because the expectation value (\ref{ci}),
    $  \langle \, \Psi (0) \vert  \, \phi({{\boldsymbol x}'}, t') \, \vert \, \Psi (0)  \, \rangle $,
   is almost zero in the region $ \vert {\boldsymbol x}' \vert > R $.
Thus, we are able to understand physically the behavior of the condensate represented in Fig.1-3.

The shape of the condensate 
   $  \langle \,  \Psi (t) \vert  \, \phi({{\boldsymbol x}}, t) \, \vert \, \Psi (t)  \, \rangle $,
   Eq.(\ref{da}), ought to be found at each time to exhibit the space-time evolution of it.
The space integral in Eq.(\ref{da}) is 
\begin{eqnarray}
   & & \int_{-\infty}^{\infty} d^3 {{\boldsymbol x}'} \, \Delta ( \vert {\boldsymbol x} - {\boldsymbol x}' \vert, t-t' ) \cdot
        \exp \left( - \frac{ 3 \vert {\boldsymbol x}' \vert^2 }{2 R^2}  \right)     \nonumber \\
  &=& \exp \left( - \frac{ 3 \vert {\boldsymbol x} \vert^2 }{2 R^2}  \right)
      \int_{0}^{\infty} 4 \pi y^2 d y \, \Delta ( y , t-t' ) \cdot 
      \exp \left( - \frac{ 3 y^2 }{2 R^2}  \right)  \cdot \left( \frac{2 R^2}{ 6 y \vert {\boldsymbol x} \vert } \right) 
       \sinh \left(  \frac{ 6 y \vert {\boldsymbol x} \vert }{2 R^2}  \right),    \nonumber \\
  \label{dd}
\end{eqnarray}
where $y \equiv \vert {\boldsymbol x} - {\boldsymbol x}' \vert $.
Using the expansion into power series,
\begin{equation}
     \sinh \left(  \frac{ 6 y \vert {\boldsymbol x} \vert }{2 R^2}  \right)
     =   \frac{ 6 y \vert {\boldsymbol x} \vert }{2 R^2} + \frac{1}{3!}
      \left( \frac{ 6 y \vert {\boldsymbol x} \vert }{2 R^2} \right)^3
    + \cdots,
  \label{de}
\end{equation}
we get 
\begin{eqnarray}
   & & \int_{-\infty}^{\infty} d^3 {{\boldsymbol x}'} \, \Delta ( \vert {\boldsymbol x}
       - {\boldsymbol x}' \vert, t-t' ) \cdot
        \exp \left( - \frac{ 3 \vert {\boldsymbol x}' \vert^2 }{2 R^2}  \right)     \nonumber \\
  & = &  \exp \left( - \frac{ 3 \vert {\boldsymbol x} \vert^2 }{2 R^2}  \right) 
      \int_{0}^{\infty} 4 \pi y^2 d y \, \Delta ( y , t-t' ) \cdot 
      \exp \left( - \frac{ 3 y^2 }{2 R^2}  \right)     \nonumber \\
  & & + \frac{1}{3!} \left( \frac{ 3  \vert {\boldsymbol x} \vert^2 }{2 R^2} \right)
           \exp \left( - \frac{ 3 \vert {\boldsymbol x} \vert^2 }{2 R^2}  \right) 
           \int_{0}^{\infty} 4 \pi y^2 d y \, \Delta ( y , t-t' ) \cdot 
      \exp \left( - \frac{ 3 y^2 }{2 R^2}  \right)  \cdot \left( \frac{ 6 y^2 }{ R^2}  \right)     \nonumber \\
  & & + \cdots.
  \label{df}
\end{eqnarray}
Now, we neglect the terms after second term in the right  hand side of Eq.(\ref{df}) that
   are zero when ${\boldsymbol x} = {\bf 0} $,
\begin{eqnarray}
 & & \int_{-\infty}^{\infty} d^3 {{\boldsymbol x}'} \, \Delta ( \vert {\boldsymbol x} - {\boldsymbol x}' \vert, t-t' ) \cdot
        \exp \left( - \frac{ 3 \vert {\boldsymbol x}' \vert^2 }{2 R^2}  \right)     \nonumber \\
 & \approx & \exp \left( - \frac{ 3 \vert {\boldsymbol x} \vert^2 }{2 R^2}  \right) 
      \int_{0}^{\infty} 4 \pi y^2 d y \, \Delta ( y , t-t' ) \cdot 
      \exp \left( - \frac{ 3 y^2 }{2 R^2}  \right). 
  \label{dg}
\end{eqnarray}
More accurate approximation is given by taking in the higher orders, but we here approximate it
   by the first term.
With this approximation, the condensate is
\begin{eqnarray}
    & & \langle \Psi (t) \vert \, \phi({{\boldsymbol x}}, t) \, \vert \Psi (t) \rangle   \nonumber  \\
  &\approx &  \exp \left( - \frac{  \vert {\boldsymbol x} \vert^2 }{2 R^2}  \right) 
           \times v \cos \, (m \, t)   \nonumber  \\
  & &  +  \exp \left( - \frac{ \vert {\boldsymbol x} \vert^2 }{2 R^2} \times 3   \right)    \nonumber  \\
  & & \hskip0.5cm  \times  \lambda \, v^3 \int_{0}^{t} d t'   \cos^3 \, ( m \, t')
         \int_{-\infty}^{\infty} d^3 {{\boldsymbol x}'} \, \Delta ( \vert {\bf 0} - {\boldsymbol x}' \vert, t-t' )        
             \exp \left( - \frac{ 3 \vert {\boldsymbol x}' \vert^2 }{2 R^2}  \right)   + O( \lambda^2 ).  \nonumber  \\
  \label{dh}
\end{eqnarray}
The second term, as well as the first term, is a product of a factor depending on the position
    $\vert {\boldsymbol x} \vert $ and a factor depending on the time $t$.
This factorization is owing to the approximation made in Eq.(\ref{dg}).
At the limit $ \vert {\boldsymbol x} \vert \rightarrow 0 $, the expression (\ref{dh}) coincides with
   Eq.(\ref{db}), as it should be.
The space dependence of the condensate is easily seen as follows.
In the term of the order  ${\cal O}(\lambda^0)$ in the condensate (\ref{dh}), the size
   is about ${\sqrt 2} R $.
On the other hand, in the term of the order  ${\cal O}(\lambda)$ in the condensate (\ref{dh}), the size
   is about ${\sqrt 2} R /{\sqrt 3} $.
Therefore, in the early time $ 0<t \ll m R^2/2 $, the contribution of the interaction term
   ${\cal H}_{\rm I} = ( \lambda /4 ) : \phi^4 : $ is restricted in rather small region
   around the origin of the condensate (see Fig.4 and Fig.5).
\begin{figure}
   \centerline{\includegraphics[height=6cm] {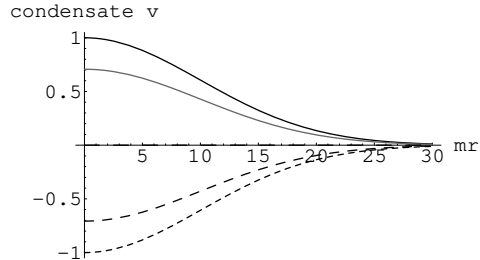}}
   \caption{ Space-time evolution of the condensate for $R_m =10$
             and $\lambda =0$ (free theory) with
             $t_m = {8 \pi \over 4}, {9 \pi \over 4}, {10 \pi \over 4},
                    {11 \pi \over 4}, {12 \pi \over 4}$, from top to bottom.}
\label{fig:4}
\end{figure}
\begin{figure}
   \centerline{\includegraphics[height=6cm] {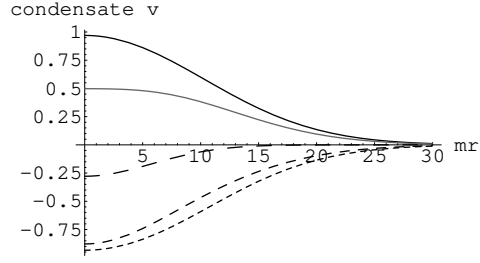}}
   \caption{ Same as Fig.4 except $\lambda =0.1$ (interaction theory).}
\label{fig:5}
\end{figure}
\subsection{Condensate when $ t < m R^2/2 $ }
In this subsection, we consider the condensate during $0< t < m R^2/2 $.
When the time $ t' < m R^2/2 $, the expectation value 
   $  \langle \, \Psi (0) \vert \, \phi({{\boldsymbol x}'}, t') \, \vert \Psi (0) \, \rangle $ 
   can not be approximated by Eq.(\ref{ci}) but by Eq.(\ref{ch}).
Since the computer has limited power, we further make an approximation in the right
   hand side of Eq.(\ref{ch}) by which numerical calculations are possible for us.
Under the condition, $R_m >>1 $ and $  t'_m < R_m^2/2 $, the following
   approximations are made,
\begin{eqnarray}
   & &  \exp \left\{ -{1 \over 2} \left( {r'_m \over R_m} \right)^2 \right\}
       \left[ 1 - {1 \over 2 !} \left( \frac{t'_m}{2 R_m^2} \right)^2
        15 \left\{ 1 -{2 \over 3} \left( {r'_m \over R_m} \right)^2 
                + {1 \over 15} \left( {r'_m \over R_m} \right)^4 \right\} \right] 
                                                                             \nonumber  \\
   & \approx & \exp \left\{ -{1 \over 2} \left( {r'_m \over R_m} \right)^2 \right\}
       \left[ 1 - {1 \over 2 !} \left( \frac{t'_m}{2 R_m^2} \right)^2
        15      \right],                  
  \label{di}
\end{eqnarray}
and
\begin{equation}
     \exp \left\{ -{1 \over 2} \left( {r'_m \over R_m} \right)^2 \right\}
    \left[ 1 - {1 \over 3} \left( {r'_m \over R_m} \right)^2  \right]    
   \approx  \exp \left\{ -{1 \over 2} \left( {r'_m \over R_m} \right)^2 \right\},
   \label{dj}
\end{equation}
which are exact at the origin $ r'_m = 0 $.
These approximations lead to
\begin{equation}
   \langle \Psi (0) \vert  \, \phi({{\boldsymbol x}'}, t') \, \vert \, \Psi (0) \rangle  
   \approx  v \exp \left\{ -{1 \over 2} \left( {r'_m \over R_m} \right)^2 \right\} F(t'),
  \label{dk}
\end{equation}
where
\begin{equation}
  F(t') \equiv  \left[  \cos(t'_m) \left\{ 1 - {15 \over 2!} \left( \frac{t'_m}{2 R_m^2} \right)^2 \right\}
      -\sin(t'_m) 
            \left\{ {3 \over 1!} \left( \frac{t'_m}{2 R_m^2} \right)
              \right\} \right].
  \label{dl}
\end{equation}
Thus, the expectation value
   $  \langle \, \Psi (0) \vert  \, \phi({{\boldsymbol x}'}, t') \, \vert \, \Psi (0)  \, \rangle $
   can be expressed as a product of a factor depending on the position $r'_m$ and a factor
    $ F(t') $ depending on the time $t'$, through which the calculation of the condensate becomes
   simple.
Substituting Eq.(\ref{dk}) into Eq.(\ref{bh}), one obtains
\begin{eqnarray}
  & & \langle \Psi (t) \vert \, \phi({{\boldsymbol x}}, t) \, \vert \Psi (t) \rangle   \nonumber  \\
  & \approx &  v \exp \left( - \frac{  \vert {\boldsymbol x} \vert^2 }{2 R^2}  \right)  F(t)   \nonumber  \\
  & &  +   \lambda \, v^3 \int_{0}^{t} d t'   F(t')^3
         \int_{-\infty}^{\infty} d^3 {{\boldsymbol x}'} \, \Delta ( \vert {\boldsymbol x} - {\boldsymbol x}' \vert, t-t' )        
             \exp \left( - \frac{ 3 \vert {\boldsymbol x}' \vert^2 }{2 R^2}  \right)   + {\cal O}( \lambda^2 ). 
  \label{dm}
\end{eqnarray}
With the help of the approximation (\ref{dg}) that has been also used in deriving the
   expression (\ref{dh}), we have for $ t< m R^2/2 $,
\begin{eqnarray}
    & & \langle \Psi (t) \vert \, \phi({{\boldsymbol x}}, t) \, \vert \Psi (t) \rangle   \nonumber  \\
  &\approx &  v \exp \left( - \frac{  \vert {\boldsymbol x} \vert^2 }{2 R^2}  \right) 
            F(t)   \nonumber  \\
  & &  +  \exp \left( - \frac{ \vert {\boldsymbol x} \vert^2 }{2 R^2}  \times 3  \right)    \nonumber  \\
  & & \hskip0.5cm  \times  \lambda \, v^3 \int_{0}^{t} d t'   F(t')^3 
         \int_{-\infty}^{\infty} d^3 {{\boldsymbol x}'} \, \Delta ( \vert {\bf 0} - {\boldsymbol x}' \vert, t-t' )        
             \exp \left( - \frac{ 3 \vert {\boldsymbol x}' \vert^2 }{2 R^2}  \right)   + O( \lambda^2 ).  \nonumber  \\
  \label{dn}
\end{eqnarray}
For early times $ t \ll m R^2/2 $, this expression coincides with Eq.(\ref{dh}),
   as it should be.

In the previous subsection, to find the space-time evolution of the condensate for
   early times $ t \ll m R^2/2 $, we have done the numerical calculation of the condensate
   (\ref{dh}) at the point $ {\boldsymbol x} = {\bf 0} $ (see Fig.1-3).
Here, we shall again carry out the numerical calculation of the condensate 
   at the point $ {\boldsymbol x} = {\bf 0} $ for $ t < m R^2/2 $, in which the expression (\ref{dn}) should
   be used instead of Eq.(\ref{dh}).
The parameters taken in the numerical calculations are the same with those adopted in the previous
   subsection, i.e., $ R_m = 10^4, 15, 10 $.
The results of the numerical calculations are shown in Fig.6, Fig.7, and Fig.8,
   from which one can see again that
   the growth of the condensate's amplitude at $ {\boldsymbol x} = {\bf 0} $ is clearly slowed down
   when its initial size
   $R_m$ becomes smaller, as was also observed in Fig.1-3.
%%%%%%%%%%%%%%%%%%%%%%%%%%%%%%%%%%%%%%%%%%%%%%%
\begin{figure}
   \centerline{\includegraphics[height=6cm] {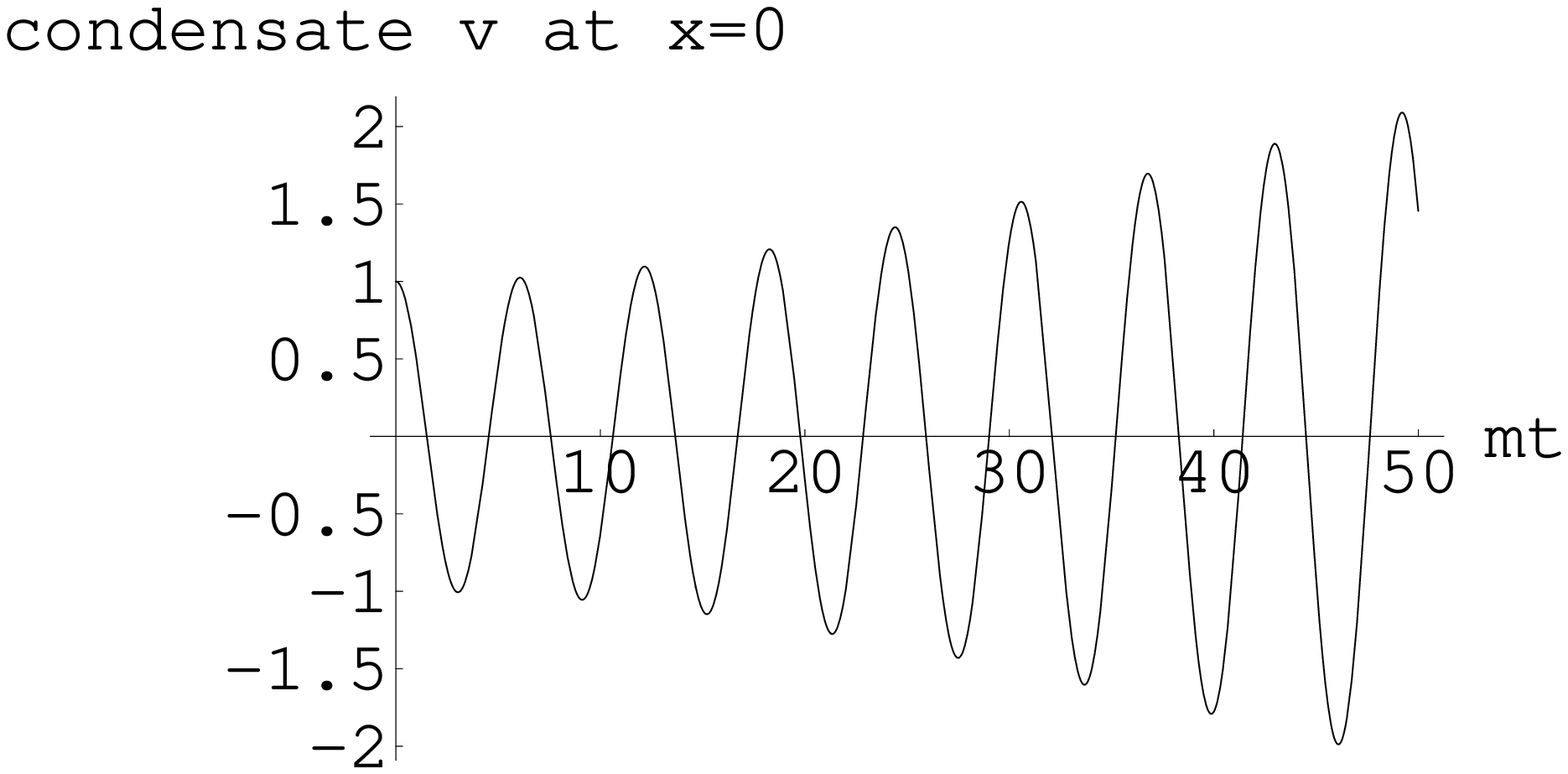}}
   \caption{ Time evolution of the condensate at ${\boldsymbol x}={\bf 0}$ 
             for $R_m =10^4$ when $0<t_m<50$.}
\label{fig:6}
\end{figure}
\begin{figure}
   \centerline{\includegraphics[height=6cm] {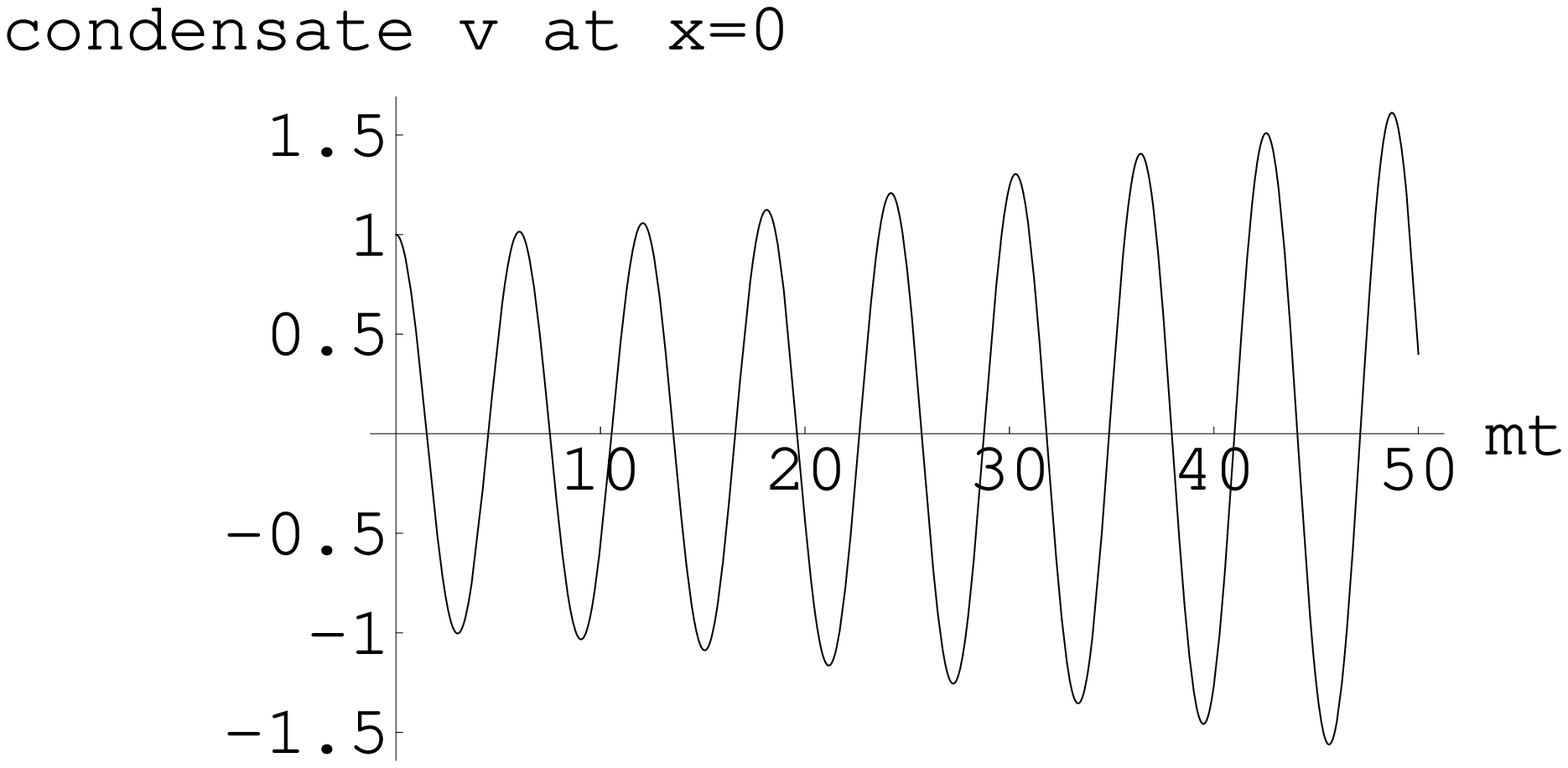}}
   \caption{ Time evolution of the condensate at ${\boldsymbol x}={\bf 0}$ 
             for $R_m =15$ when $0<t_m<50$.}
\label{fig:7}
\end{figure}
\begin{figure}
   \centerline{\includegraphics[height=6cm] {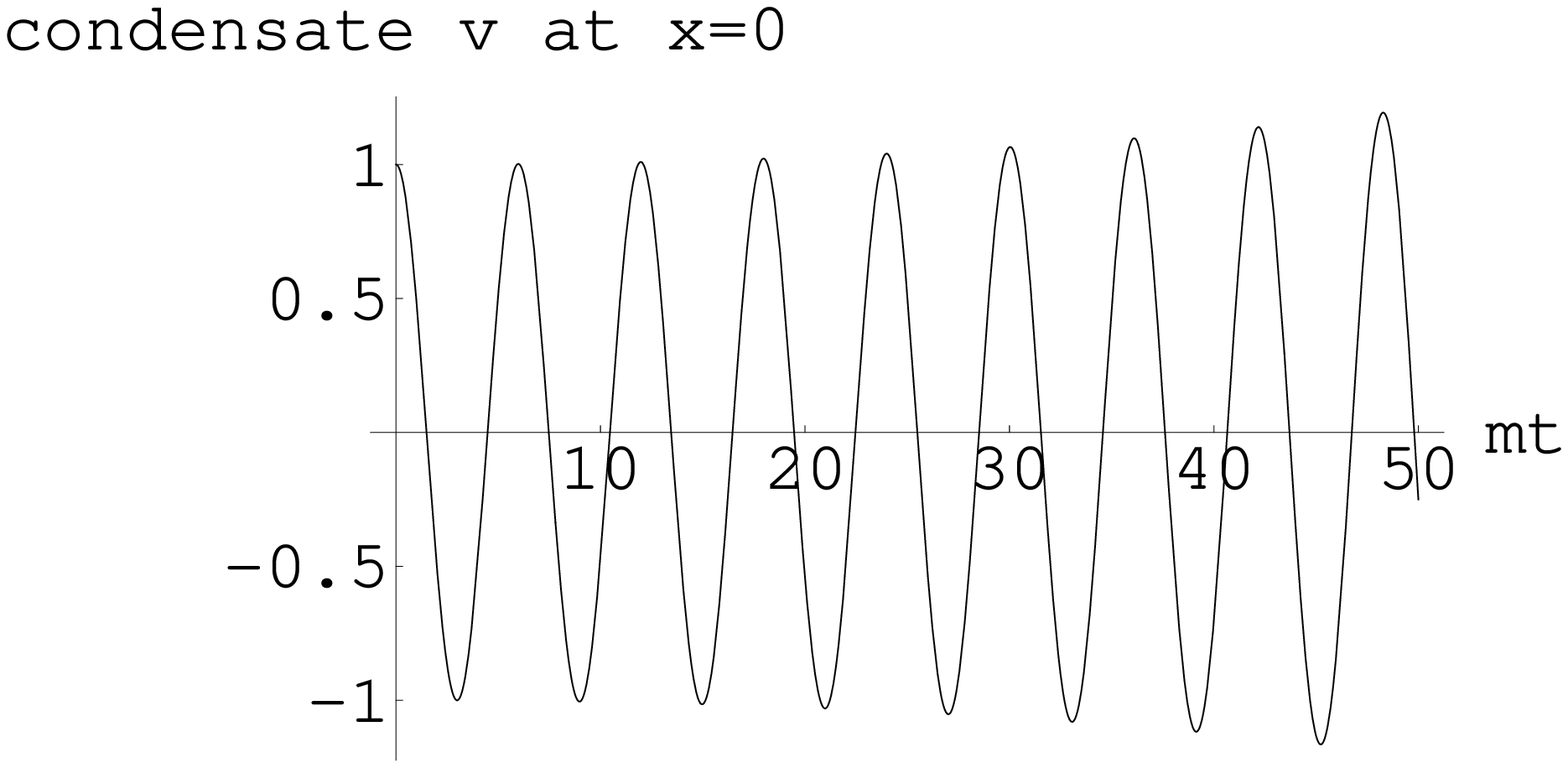}}
   \caption{ Time evolution of the condensate at ${\boldsymbol x}={\bf 0}$ 
             for $R_m =10$ when $0<t_m<50$.}
\label{fig:8}
\end{figure}
%%%%%%%%%%%%%%%%%%%%%%%%%%%%%%%%%%%%%%%%%%%%%%%%%%%%%%%%%%
The space dependence of the condensate (\ref{dn}) for $ t < m R^2/2 $ is similar to that of
   the condensate (\ref{dh}) for $ t \ll m R^2/2 $.
Namely, the space dependent factors 
   $ \exp \left( -  \vert {\boldsymbol x} \vert^2 / 2 R^2  \right)  $ and
   $ \exp \left( -  3 \times \vert {\boldsymbol x} \vert^2 / 2 R^2  \right)  $
   are the same. 
Consequently, one can say that, even for $ t < m R^2/2 $, the contribution of the interaction
   term ${\cal H}_{\rm I} $ is restricted in a small region around the origin of the condensate.

Thus, it is shown that the space-time evolution of the condensate for $ t < m R^2/2 $ resembles
   that of the condensate for $ t \ll m R^2/2 $ in behavior.
\newpage
\section{Conclusion}
We studied the space-time evolution of the condensate in the interaction representation.
The model is the  $ \lambda \phi^4 $ theory and the initial physical state is taken to be
   a coherent state.
In the case of the homogeneous condensate, the amplitude of the condensate grows in proportion to 
   the time, which implies that the perturbative calculation breaks down for large time.
On the other hand, in the case of the inhomogeneous condensate with the finite size
   $\sim R$ at the initial time $t=0$, the space-time evolution is calculated for 
   $ 0<t < m \, R^2/2 $.
The growth of the condensate's amplitude at ${\boldsymbol x}={\bf 0}$ is slowed down when its initial size
   $R$ becomes smaller.
The condensate maintains its initial size $R$ and the interaction term 
   ${\cal H}_{\rm I} = ( \lambda /4 ) : \phi^4 : $ has an effect in the
   vicinity of the origin of the condensate.
When the time $t$ exceeds the value $m R^2/2$, the condensate spreads.
The reasons for the spread are twofold;
   first, the condensate behaves as a wave due to the initial condition taken to be a 
   coherent state and
   second, many different modes are superposed to locate the condensate at the initial time.
Thus, it is recognized that the space-time evolution of the inhomogeneous condensate is
   so different from that of the homogeneous one.

In the case of the homogeneous condensate in the  $ \lambda \phi^4 $ theory,
   the perturbative calculation breaks down for large time as was shown in
   the section 2.
However, if the condensate has a finite size $R$ at the initial time, one can expect 
   the perturbative method is applicable even for late times.
Because, the growth of the amplitude of the inhomogeneous condensate is slowed down
   compared with that of the homogeneous one for $t < m \, R^2/2 $, and the behavior of
   the expectation value 
   $  \langle \, \Psi (0) \vert  \, \phi({{\boldsymbol x}'}, t') \, \vert \, \Psi (0)  \, \rangle $
   of the inhomogeneous condensate is different from that of the homogeneous one.

Our specific analysis is indebted to the interaction representation in which
   the field operator obeys the free equation of motion.
Since the properties of the free field operator are well known, we utilize them
   such as the Lorentz invariant function $ \Delta ( \vert {\boldsymbol x} \vert, t )  $.
Moreover, the particle picture can be obtained definitely in that representation.
The QCD or the sigma model which is an effective theory of QCD has a strong coupling
   constant, and unfortunately the interaction representation can not be applied to these theories.
However, studying the model with weak coupling constant by use of the interaction representation 
   would be helpful in order to understand more physically the space-time evolution of the condensate
   or the quantum particle emission accompanied by the decay of the condensate.
An example of such a model is the linear sigma model with the weak coupling constant 
   ($\lambda<1$).
Assuming the $\pi$ field condensation \cite{rf:Mae} with a finite size, one may be able to study the 
   $\pi$ or $\sigma$ quantum particle emissions accompanied by the decay of the condensate,
   where the quantum particle picture is very clear despite the inhomogeneous condensate.
\section*{Acknowledgements}
The author thanks the Yukawa Institute for Theoretical Physics at Kyoto University,
    where this work was initiated during the
   YITP-W-03-10 on "Thermal Quantum Field Theories and Their Applications". 
\appendix
\section{} %Empty argument \section{} yields `Appendix'. 
%
%\section{Second Appendix}
%
The Lorentz invariant function defined in Eq.(\ref{bd}) is expressed explicitly as
\begin{equation}
  \Delta  ( \vert {\boldsymbol x}  \vert, t )
 = - \frac{1}{2 \pi} \epsilon ( t) \left[  \delta ( t^2 - \vert {\boldsymbol x} \vert^2 ) 
      - \frac{m^2}{2} \, \theta (  t^2 - \vert{\boldsymbol x} \vert^2 ) \,
        \frac{ J_1 ( m \sqrt{ t^2 - \vert{\boldsymbol x} \vert^2 } ) }
             {  m \sqrt{ t^2 -\vert {\boldsymbol x} \vert^2 } }  \right], 
  \label{fa}
\end{equation}
where $J_1(z) $ is the Bessel function of the first kind.
For $t-t'>0$,
\begin{eqnarray}
  & &  \int_{-\infty}^{\infty} d^3 {{\boldsymbol x}'} \,
       \Delta (  \vert {\bf 0} - {\boldsymbol x}' \vert, t-t' ) \cdot
        \exp \left( - \frac{ 3 \vert {\boldsymbol x}' \vert^2 }{2 R^2}  \right)     \nonumber \\
  &=&  \int_{0}^{\infty} 4 \pi y^2 d y \, \Delta ( y , t-t' ) \cdot 
      \exp \left( - \frac{ 3 y^2 }{2 R^2}  \right)    \nonumber \\
  &=&  -(t-t' ) \exp \left(-\frac{3(t-t')^2}{2 R^2} \right)         \nonumber \\
   & & \hskip1cm   + m^2 \int_{0}^{t-t'} d y \, y^2 \,
          \frac{  J_1( m \sqrt{ (t-t')^2 - y^2 } \, ) }{m \sqrt{ (t-t')^2 - y^2 } }
          \exp \left(- \frac{ 3 y^2 }{2 R^2}  \right).
  \label{fb}
\end{eqnarray}
Especially when $R=\infty$, one obtains
\begin{eqnarray}
  & &  \int_{-\infty}^{\infty} d^3 {{\boldsymbol x}'} \,
         \Delta ( \vert {\bf 0} - {\boldsymbol x}' \vert, t-t' )        \nonumber \\
  &=&  -(t-t' ) 
           + {1 \over m}  \int_{0}^{ m \, (t-t') } d s \,
             \sqrt{ m^2 (t-t')^2 - s^2 } \, J_1 (s)                    \nonumber \\
  &=&  -{1 \over m} \sin \{m (t-t' ) \},
  \label{fc}
\end{eqnarray}
where the following integral formula has been used,
\begin{equation}
   \int_{0}^{ \alpha } d s \, \sqrt{ \alpha^2 -s^2 } \, J_1 (s) 
       = \alpha - \sin \alpha.
  \label{fd}
\end{equation}
%

%%%%%  up to here 2004.11.17  %%%%%%%%%%%%%%%%%%%%%%%%%%%%%%%%%%%%%%%%%%%%%%%%%%%%%%
%%%%%%%%%%%%%%%%%%%%%%%%%%%%%%%%%%%%%%%%%%%%%%%%%%%%%%%%%%%%%%%%%%%%%%%%%%%%%%%%%%%%%%%%%%
% 

%
\end{document}